\documentclass[journal,twoside]{IEEEtran}
\usepackage{cite}
\usepackage{graphicx}
\usepackage{amsmath}
\usepackage[normalem]{ulem}
\usepackage{float}
\usepackage{subfig}
\usepackage{ulem}

\newcommand{\dif}{\mathrm{d}} 
\newcommand{\me}{\mathrm{e}} 
\newcommand{\mj}{\mathrm{j}} 
\newcommand{\nm}{\mathrm{nm}} 
\newcommand{\um}{\mathrm{{\mu}m}} 
\newcommand{\mm}{\mathrm{mm}} 

\begin{document}
\title{Highly Efficient Boundary Element Analysis of Whispering Gallery Microcavities
\author{Leyuan~Pan and Tao~Lu}
\thanks{The authors are with the Department of Electrical and Computer Engineering, University of Victoria, Victoria, BC, V8P 5C2 Canada (email: taolu@ece.uvic.ca).}}
\maketitle
\begin{abstract}
We demonstrate that the efficiency of the boundary element whispering gallery microcavity analysis can be improved by orders of magnitude with the inclusion of Fresnel approximation. Using this formulation, simulation of a microdisk with wave-number-radius product as large as $kR\approx8,000$ was demonstrated in contrast to a previous record of $kR\approx100$. In addition to its high accuracy on computing the modal field distribution and resonance wavelength, this method yields a relative error of $10\%$ in calculating the quality factor as high as $10^{11}$ through a direct root searching method where the conventional boundary element method failed to achieve. Finally, quadrupole shaped cavities and double disks as large as $100~\um$ in diameter were modeled by employing as few as $512$ boundary elements whilst the simulation of such large cavities using conventional boundary element method were not reported previously.
\end{abstract}
\section{Introduction}
Ultra-high Quality factor (Q) whispering gallery microcavities~\cite{Vahala2003,Armani2003,Gorodetsky1996Ultimate,Vollmer2008,Lu12042011,Dantham_Label,Lu_Min_Narrow_Laser,Li:14,PhysRevLett.109.233901,HyunKim2013} are at the research frontiers of nanosensing, frequency comb generation, nonlinear optics, ultra-narrow linewidth laser and wave chaos. Guiding the related researches requires highly efficient numerical techniques for precise modelling of the cavity and its wave propagation behaviour. Among a variety of numeric techniques reported\cite{Li2012Micro,Du2013a,Shirazi:13},  boundary element method (BEM) has a proven record on its accuracy and efficiency both in traditional waveguide\cite{book_Brebbia_Boundary,Yang_Boundary,Lu:02,Lu2003,pmd_Lu_Comparative} and whispering gallery microcavity analysis\cite{Kagami_Application,Wiersig2003,Zou2009Accurately}. BEM employs the Green's theorem to associate the electro-magnetic field in an isotropic and homogeneous medium to that of its boundaries. Consequently one only needs to compute the field at the media boundaries before computing it elsewhere. This reduces the number of elements involved in the computation by one dimension compared to the finite element or finite difference methods where a full discretization of the whole optical structure is required. Albeit its relative higher efficiency, the conventional boundary element analysis of whispering gallery microcavities is impeded by the relative large number of elements required for discretizing cavity boundaries.  As an illustration, a two dimensional modelling of $1,550$ nm wavelength light propagating in a $4$-mm-diameter silica microdisk typically used in frequency comb generation~\cite{PhysRevLett.109.233901} requires the discretization of the microdisk boundary to a minimum of around $24,000$ elements such that the smaller-than-half-wavelength requirement of the element size is met according to the Nyquist theorem. Consequently, a scalar BEM needs to solve a $48,000{\times}48,000$ matrix for modelling such cavities. This well exceeds the computational calibre of a conventional desktop computer.

In this paper, we implement a slow wave, also known as Fresnel approximation in boundary element whispering gallery microcavity analysis in analogy to that developed in~\cite{Lu_Master} for straight waveguide analysis. Using this technique, we calculate the slow varying field envelope at the cavity boundaries instead of the rapid varying boundary field calculated by the conventional boundary element method formulation. Therefore, the discretized element size can be orders of magnitude larger than that required by the conventional method without loss of accuracy.

\section{Formulation}
\begin{figure}[htbp]
\centering\includegraphics[width=\linewidth]{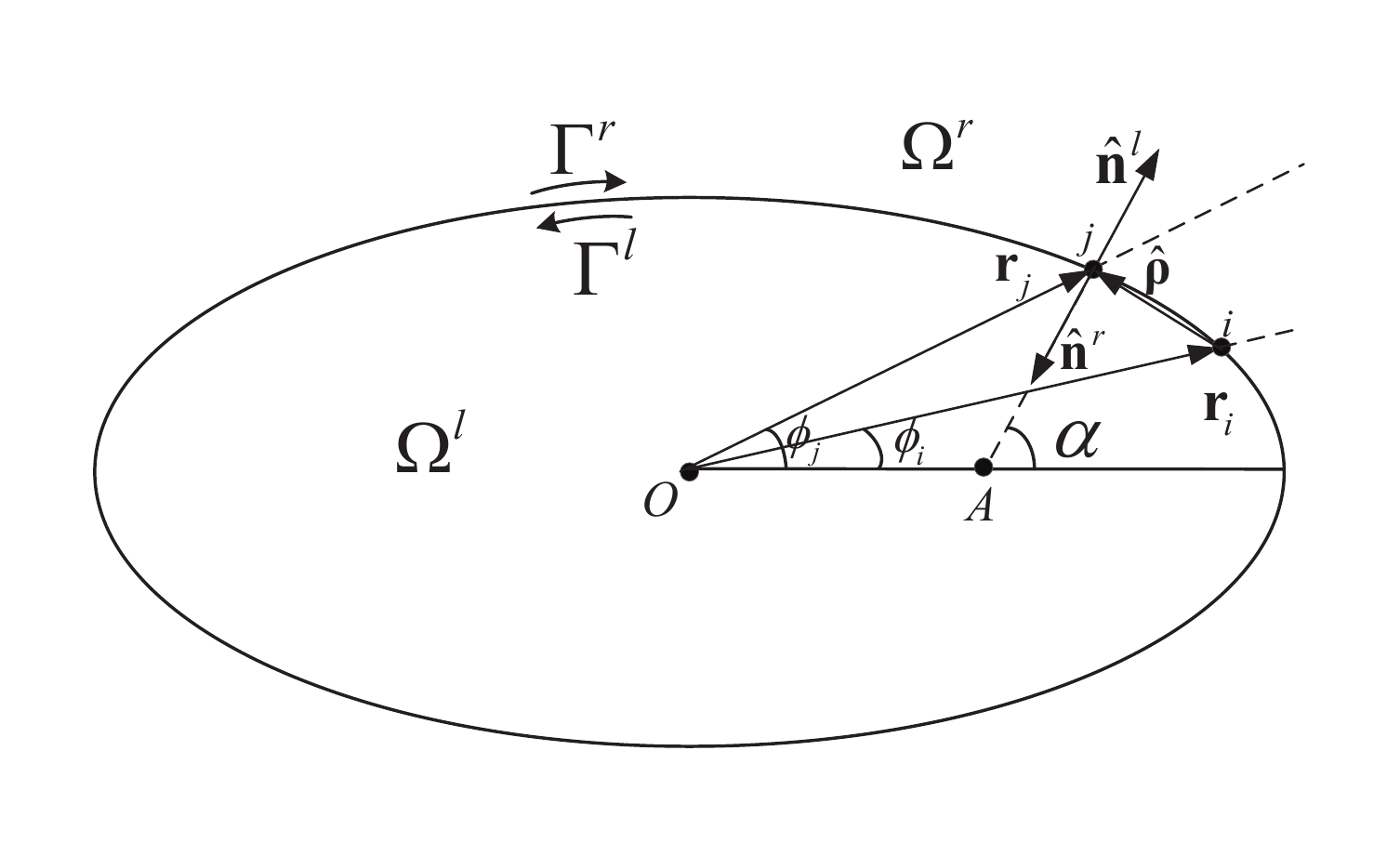}
\caption{Top view of a Whispering Gallery Microcavity. }
\label{fig:CircularDomain2D}
\end{figure}

\begin{figure*}[htbp]
\centering
\subfloat[]{
\label{fig:LinearSearchR5um}
\begin{minipage}[t]{0.32\linewidth}
\centering\includegraphics[width=\linewidth]{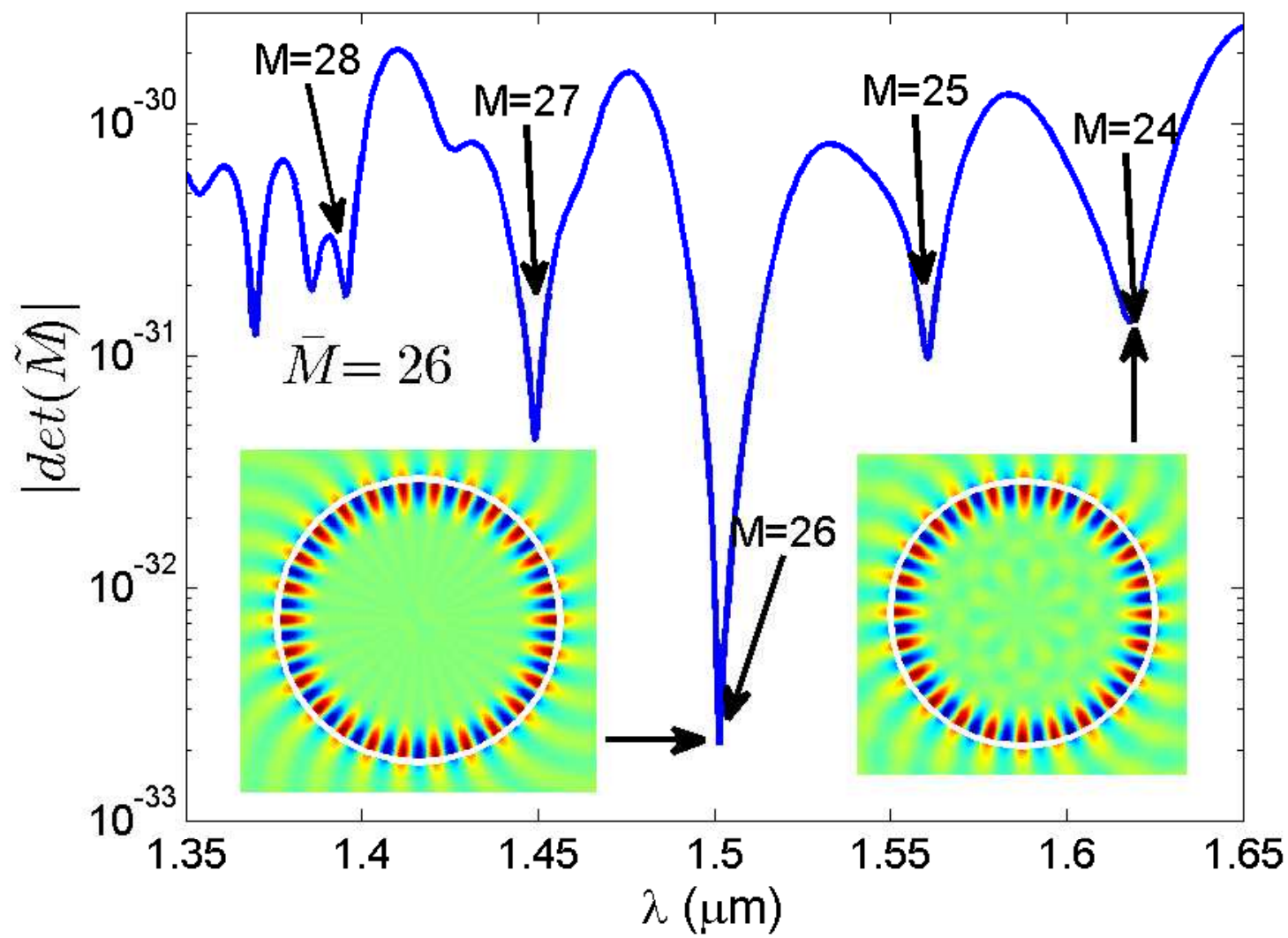}
\end{minipage}}
\vspace{-0.15em}
\subfloat[]{
\label{fig:BoundFieldsWaterFall}
\begin{minipage}[t]{0.32\linewidth}
\centering\includegraphics[width=\linewidth]{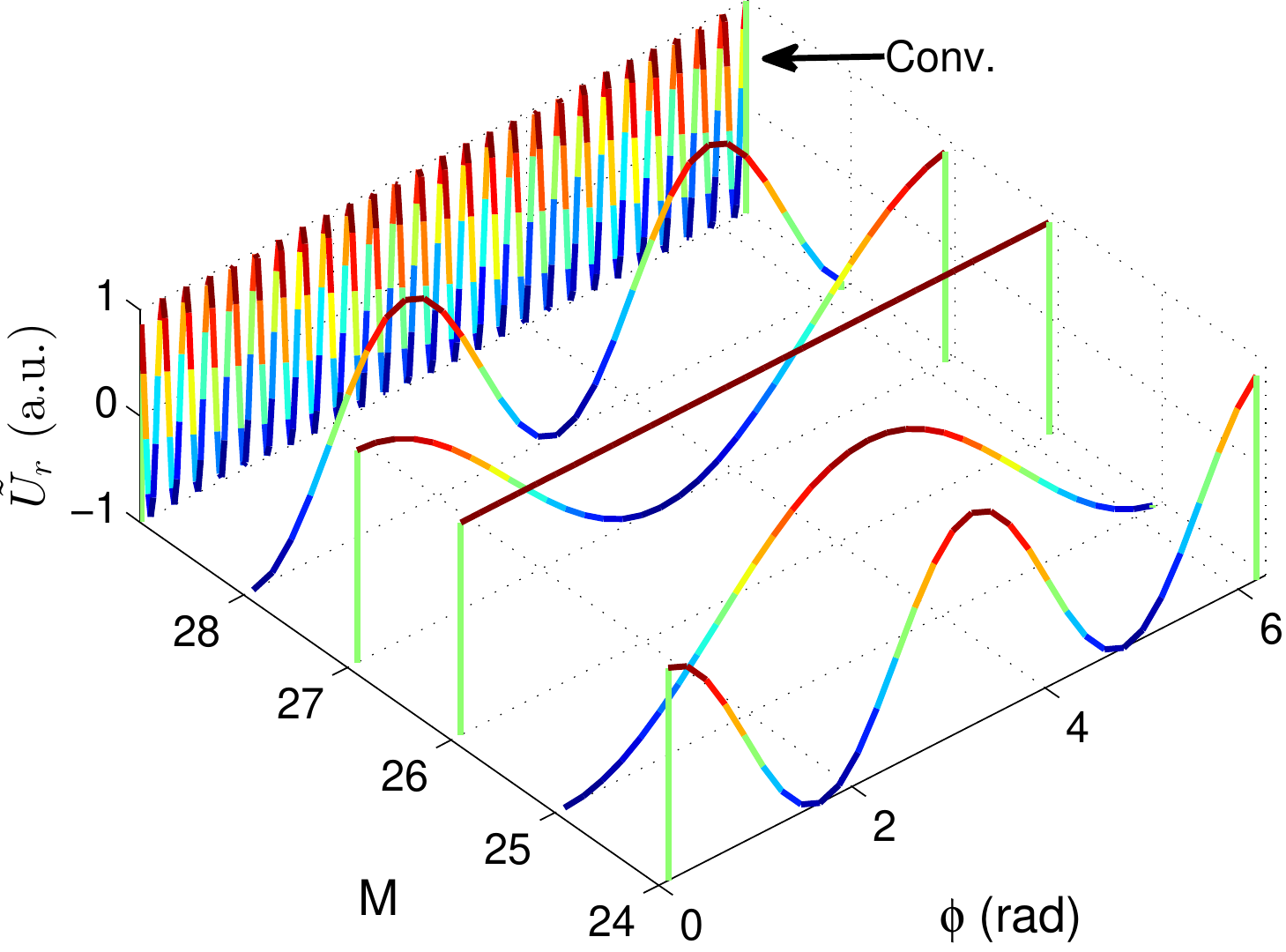}
\end{minipage}}
\vspace{-0.15em}
\subfloat[]{
\label{fig:CompareFresnelNonFresnelGauss32R5umMOrder25}
\begin{minipage}[t]{0.32\linewidth}
\centering\includegraphics[width=\linewidth]{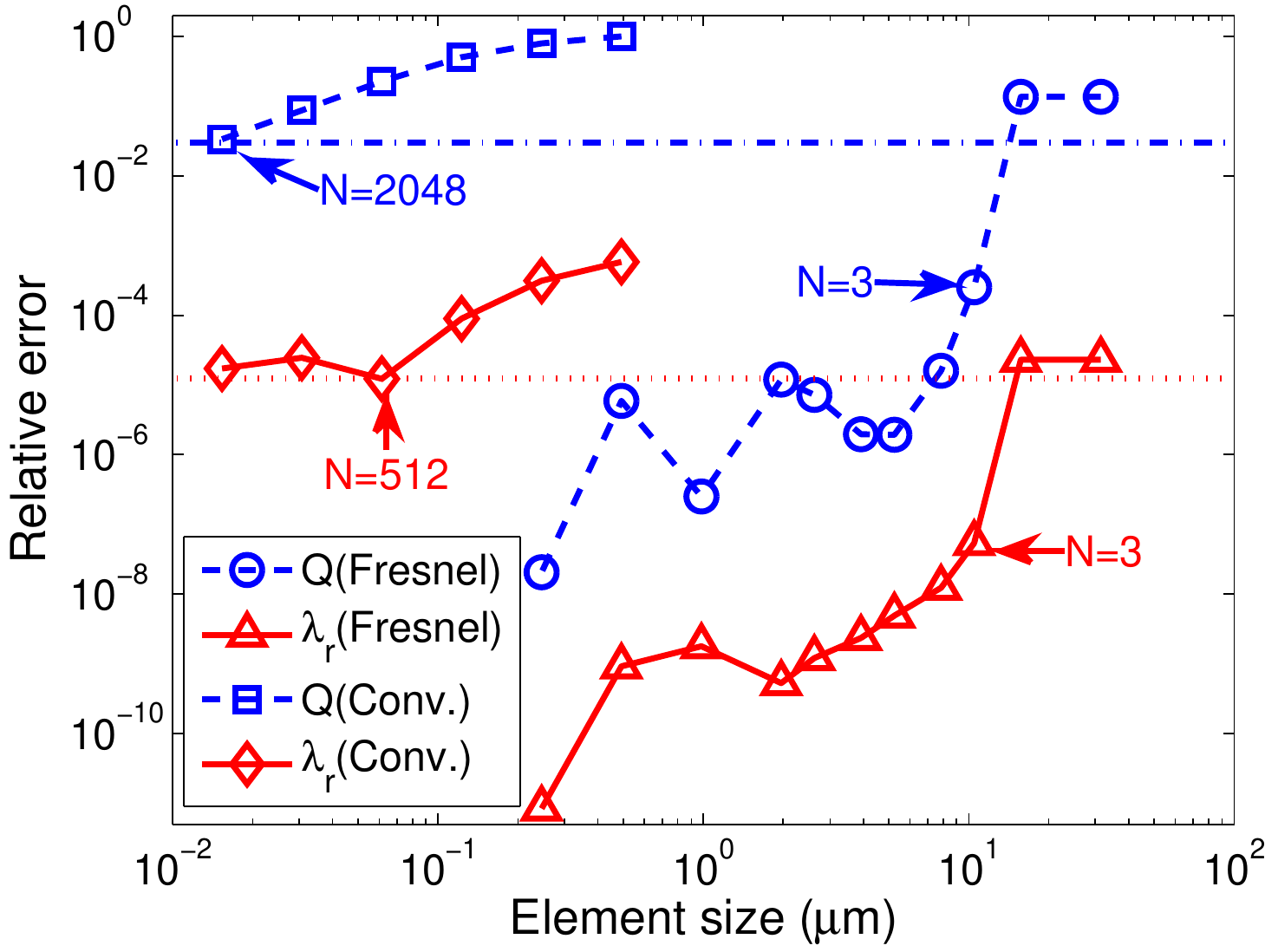}
\end{minipage}}
\caption{(a) The magnitude of the determinant of $\mathbf{\tilde M}$ vs. wavelength; (b) Boundary field envelope ${\tilde U}$ of the $24^{th}$ to $28^{th}$ azimuthal modes calculated by Fresnel approximation $({\bar M}=26)$ vs. the $26^{th}$ order field $U$ calculated by conventional method; (c) Relative error of resonance wavelength and quality factor as a function of the boundary element size using Fresnel approximation and conventional method.}
\end{figure*}

According to Green's Theorem\cite{Jackson1975}, in a two dimensional source-free, isotropic, piece-wise homogeneous medium of refractive index ${\tilde n}$, the electro-magnetic (E-M) field $U$ at a wavelength $\lambda$ satisfies the path integral\cite{Yang_Boundary,Lu:02}
\begin{equation}
    \label{eq:generalBIE-1}
    {\gamma}U({\mathbf{r}}) = - \int_{\Gamma-\Gamma_\epsilon}U(\mathbf{r^\prime}) \frac{\partial G(\mathbf{r^\prime}, \mathbf{r})}{{\partial} n^\prime}\dif{l^\prime} + \int_{\Gamma-\Gamma_\epsilon} \frac{\partial U(\mathbf{r^\prime})}{{\partial}n^\prime}G(\mathbf{r^\prime}, \mathbf{r})\dif{l^\prime}
\end{equation}
where following the definition in~\cite{Lu:02} (cf. Fig.~\ref{fig:CircularDomain2D}), $\mathbf{r'}$ is the position vector of the field point at medium boundaries and $\mathbf{r}$ is that of the observation point. $\gamma$ is a constant with unity value if $\mathbf{r}$ is inside the medium and $1/2$ if it is located at a smooth boundary. $\Gamma$ represents the complete set of boundaries enclosing the medium and $\Gamma_\epsilon$ is the infinitesimal boundary at the observation point if it is located at one of the boundaries. $\frac{\partial f(\mathbf{r})}{{\partial}n}=\mathbf{\hat n}{\cdot}{\nabla}f(\mathbf{r})$ is the normal derivative of a functional $f(\mathbf{r})$ at a boundary point $\mathbf{r}$ where $\mathbf{\hat n}$ is the unit normal vector of the boundary at $\mathbf{r}$ pointing outward from the medium. Under the scalar approximation, $U(\mathbf{r^\prime})$ represents any component of the field. The Green function $G(\mathbf{r^\prime}, \mathbf{r})=\frac{1}{4j}H^{(2)}_0({\tilde n}(\mathbf{r^\prime})k_0|\mathbf{r^\prime}-\mathbf{r}|)$ has the form of the zeroth order Hankel function of the second kind $H^{(2)}_0$ in two dimensional case with $k_0=2{\pi}/\lambda$ as the wave number in free space.

In a whispering gallery microcavity, the E-M field of a $M^{th}$ order azimuthal mode can be represented in a cylindrical coordinate system $\mathbf{r}\equiv(\rho,\phi)$ according to
\begin{equation}
    \label{eq:envelope-bie-1}
    U(\rho,\phi) = \tilde{U}(\rho,\phi)e^{j{\bar M}\phi}
\end{equation}
in the case where ${\bar M}$ is a value close to $M$, $\tilde{U}(\rho,\phi)$ becomes slow varying along azimuthal direction compared to $U(\rho,\phi)$. Note that for an ideal whispering gallery microcavity when we select ${\bar M}=M$, the field envelope $\tilde{U}(\rho,\phi)$ becomes constant.

Discretizing the media boundaries into $N$ elements whose centers are located at $({\bf r_1},{\bf r_2}\ldots{\bf r_N})$ and substituting $U({\bf r_i})$ in Eq.~\eqref{eq:generalBIE-1} with $\tilde{U}({\bf r_i})$ according to Eq.~\eqref{eq:envelope-bie-1}, noting that $\tilde{U}({\bf r_i})$ simultaneously satisfies Eq.~\eqref{eq:generalBIE-1} defined in both left and right sides of the boundary, we obtain $2N$ linear equations with $2N$ unknowns defined by two vectors $ \mathbf{\tilde{U}}=\left(\tilde{U}({\bf r_1}),\tilde{U}({\bf r_2}),\cdots,\tilde{U}({\bf r_N})\right)^\mathrm{T}$ and $\mathbf{{\tilde U}^\prime}=\left(\frac{\partial\tilde{U}({\bf r_1})}{\partial{n}},\frac{\partial\tilde{U}({\bf r_2})}{\partial{n}},\cdots,\frac{\partial\tilde{U}({\bf r_N})}{\partial{n}}\right)^\mathrm{T}$ as
\begin{equation}
    \label{eq:envelope-bie-3}
    \begin{split}
        \frac{1}{2}\tilde{U}(\mathbf{r}_i)\me^{\mj\bar{M}\phi_i} = &- \sum_{j} \tilde{U}(\mathbf{r}_j) \int_{\Gamma_j^{l,r}}\me^{\mj{\bar M}\phi_j}\left[\frac{\partial G^{l,r}(\mathbf{r}_j, \mathbf{r}_i)}{\partial n^{l,r}}\right. \\
        &- \left.\frac{\mj{\bar M}\sin(\alpha^{l,r}-\phi_j)G^{l,r}(\mathbf{r}_j, \mathbf{r}_i)}{r_j}\right]\dif\Gamma \\
        &+ \sum_j\frac{\partial\tilde{U}(\mathbf{r}_j)}{\partial n^{l,r}}\int_{\Gamma_j^{l,r}}\me^{\mj{\bar M}\phi_j}G^{l,r}(\mathbf{r}_j, \mathbf{r}_i)\dif\Gamma
    \end{split}
\end{equation}
where $\alpha^{l,r}$ is the angle of the unit normal vector $\mathbf{\hat n}^{l,r}$, $\mathbf{r}_i$ and $\mathbf{r}_j$ denotes the coordinates of the element center on $\Gamma_i$ and $\Gamma_j$ as illustrated in Fig.~\ref{fig:CircularDomain2D}.

For simplicity, Eq.~\eqref{eq:envelope-bie-3} can be expressed into a matrix form
\begin{equation}
\mathbf{\tilde M}\begin{pmatrix}
\mathbf{\tilde U}\\
\mathbf{\tilde U}^\prime
\end{pmatrix}=\begin{pmatrix}
       -\mathbf{\tilde H}^l & \mathbf{\tilde G}^l \\
       -\mathbf{\tilde H}^r & -\mathbf{\tilde G}^r
\end{pmatrix}
\begin{pmatrix}
\mathbf{\tilde U}\\
\mathbf{\tilde U}^\prime
\end{pmatrix} =\mathbf{0}
\end{equation}
Here $\mathbf{\tilde H}^{l,r}$ and $\mathbf{\tilde G}^{l,r}$ are $k_0$-dependent $N{\times}N$ block sparse matrices whose $(i,j)^{th}$ matrix elements $h^{l,r}_{ij}$ and $g^{l,r}_{ij}$ can be derived from Eq.~\eqref{eq:envelope-bie-3}. Similar to the conventional BEM, by finding a complex wave number ${\tilde k}_0$ such that the matrix determinant in Eq.~\eqref{eq:envelope-bie-3} vanishes, we obtain a whispering gallery mode with the corresponding non-zero solutions of $\mathbf{\tilde U}$ as the modal field distribution at the boundary. The real part $\lambda_r$ of the complex wavelength ${\tilde \lambda}=\lambda_r-j\lambda_i=2\pi/{\tilde k}_0$ is the resonance wavelength of this mode while the quality factor can be calculated according to $Q=\lambda_r/2\lambda_i$.

\section{Results and discussions}
To compare the efficiency of the new formulation to the conventional method, we first simulated the whispering gallery mode of a two dimensional $5$-$\um$-radius silica microdisk at a resonance wavelength around $1.55~\um$ by setting ${\bar M}=26$. In Fig.~\ref{fig:LinearSearchR5um} we plot the magnitude of $\mathbf{\tilde M}$ matrix determinant in a wavelength span between $1.35~\um$ and $1.65~\um$. In this figure, five minimum dips are located at $1.61796~\um$, $1.56047~\um$, $1.50150~\um$, $1.44920~\um$ and $1.39553~\um$, corresponding to the resonance wavelengths of $24^{th}$ to $28^{th}$ azimuthal modes. Note that the minima are equally spaced at a free spectral range of $55.6~\nm$. For further verification, we plot the field distribution at wavelengths corresponding to the $24^{th}$ and $26^{th}$ modes as the insets. To illustrate the efficiency of our implementation, the boundary field envelope of the $24^{th}$ to $28^{th}$ modes are plotted in Fig.~\ref{fig:BoundFieldsWaterFall}. As seen, the field envelope of the $26^{th}$ azimuthal mode is constant under Fresnel approximation while the field envelope varies slowly with a periodicity of $2\pi/|M-{\bar M}|$ for modes whose azimuthal order deviate from ${\bar M}$. Nevertheless, it is sufficient to employ as few as $16$ boundary elements for the modal field distribution calculation. Note at a wavelength below $1.4~\um$ spurious modes start to appear as the higher azimuthal mode from the reference ${\bar M}$ requires a finer element discretization. As a comparison, the boundary field of the $26^{th}$ mode is also plotted using the conventional method. In that case, the field amplitude oscillates rapidly with a periodicity of $2\pi/26$. Consequently, as many as $256$ elements were required to achieve comparable accuracy. Fig.~\ref{fig:CompareFresnelNonFresnelGauss32R5umMOrder25} further quantifies the comparison by displaying the relative error of the computed resonance wavelength and quality factor as a function of the boundary element size. Here, the analytical solution~\cite{Ryu2008Resonances,Zou2009Accurately} is used as a reference. With the Fresnel approximation and $128$ boundary elements, the relative error of the computed resonance wavelength drops to below $10^{-10}$ while the calculated quality factor yields an error of around $10^{-8}$. On the other hand, the conventional boundary element method yields a relative error of $10^{-3}$ in computing resonance wavelength and almost unity in computing quality factor with the same number of elements. Furthermore, to reach a relative error of below $10\%$ in computing the quality factor, the conventional method requires the discretization of the boundary into $2,048$ elements while with the Fresnel approximation as few as $3$ elements is sufficient. Finally, with $3$ elements the Fresnel approximated method reaches a relative error of around $10^{-5}$ while the conventional method requires $512$ elements for the same accuracy. Clearly the Fresnel approximated technique outperforms the conventional method by orders of magnitude.

It is worth mentioning that unlike the name suggested, the solution obtained from the Fresnel approximation is exact. The efficiency is improved by orders of magnitude as a result of extracting the mode field envelope from its otherwise rapidly varying boundary field in the same manner as down converting a baseband signal from its carrier frequency, a technique commonly used in the field of communications.
\begin{figure}[htbp]
\subfloat[]{
\label{fig:QualityFactorWavelengthVSkR}
\begin{minipage}[b]{0.9\linewidth}
\centering \includegraphics[width=\linewidth]{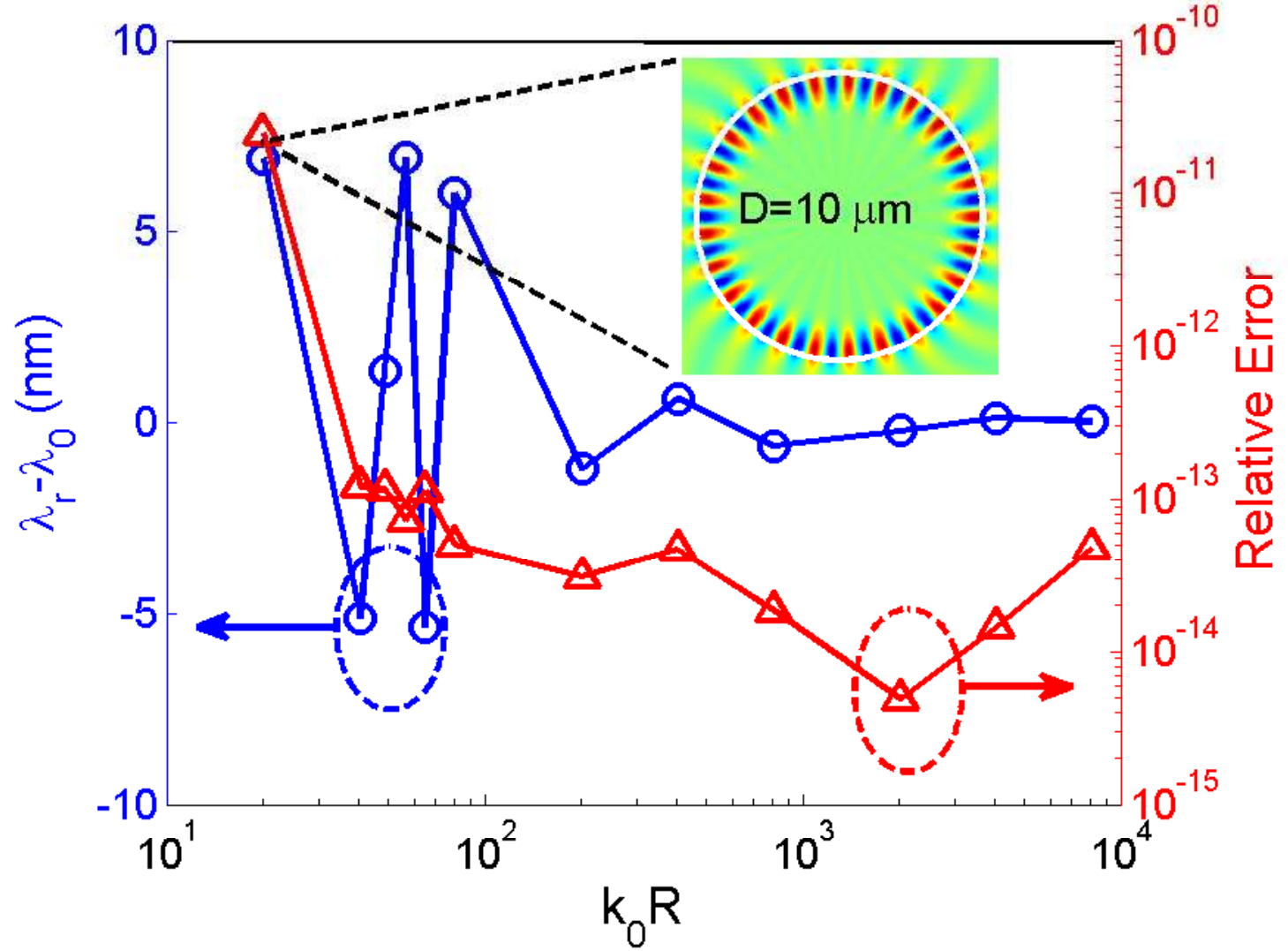}
\end{minipage}}
\vspace{-0.15em}
\subfloat[]{
\label{fig:FieldDistributionFresnel5umReal}
\begin{minipage}[b]{0.9\linewidth}
\centering \includegraphics[width=\linewidth]{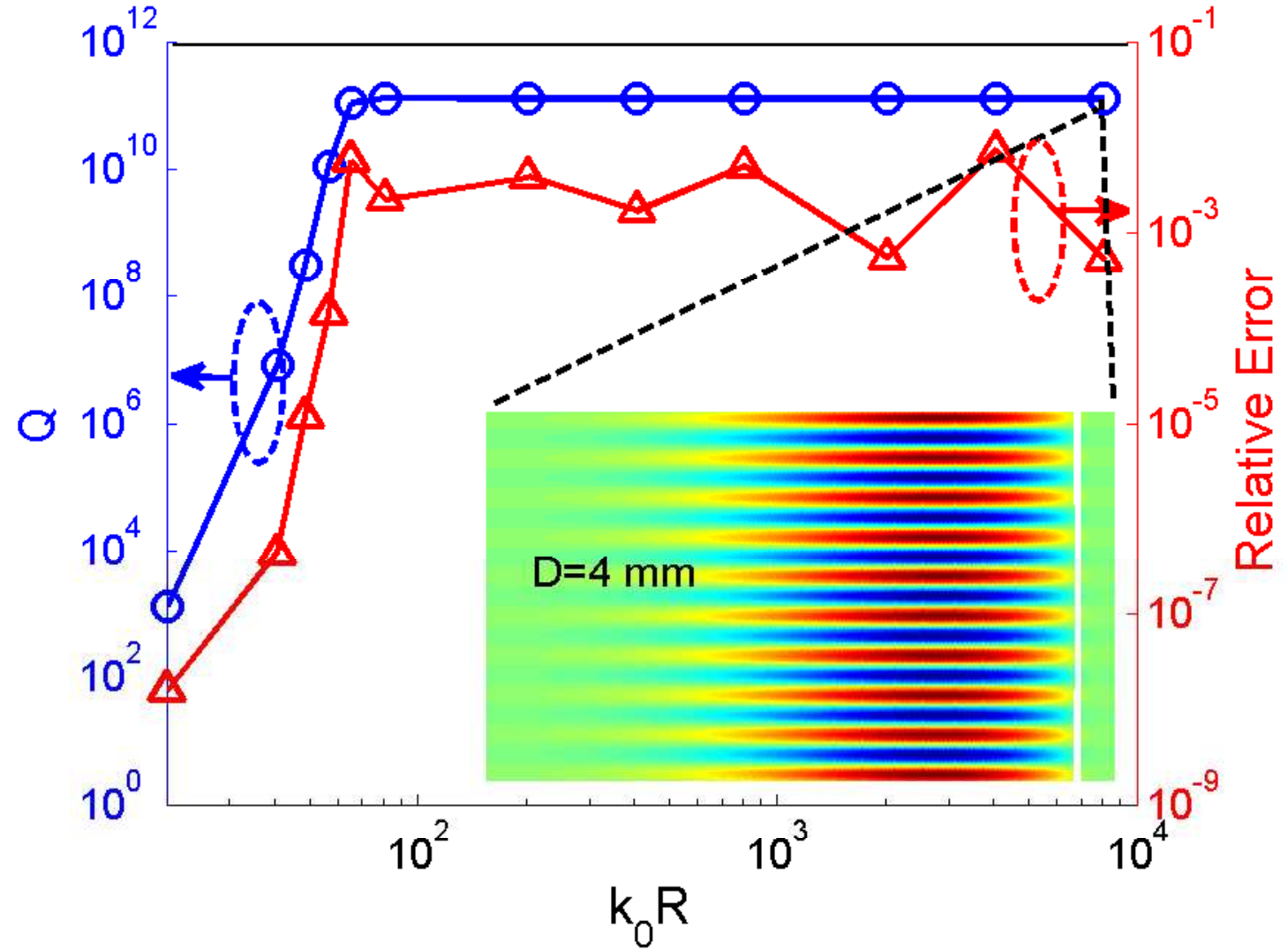}
\end{minipage}}
\vspace{-0.2em}
\caption{(a) Resonance wavelength detuning $(\lambda_0=1.55~\um)$ and (b) quality factor of silica microdisks as a function of their wave-number-radius product (blue circles in both plots). The corresponding relative errors are plotted as red triangles. The insets are the modal field distribution of microdisks with $10~\um$ and $4~\mm$ diameters. The element size is set to be around $10~\um$ in all cases.}
\label{fig:convergenceFresnel}
\end{figure}

In Fig.~\ref{fig:QualityFactorWavelengthVSkR} and Fig.~\ref{fig:FieldDistributionFresnel5umReal}, we further calculate the resonance wavelength and quality factor (both are represented as circles in the corresponding figures) as a function of wave-number-radius product $(k_0R)$. Here, the boundary element size is around $10~\um$. The relative errors are displayed as red triangles by comparing the computed results with analytical solutions. As seen, a relative error below $10^{-11}$ is achieved in resonance wavelength computation and $10^{-2}$ in quality factor. The corresponding modal field distribution of microdisks with diameter of $10~\um$ and $4~\mm$ are displayed as insets of both Fig.~\ref{fig:QualityFactorWavelengthVSkR} and Fig.~\ref{fig:FieldDistributionFresnel5umReal}. Evidentally, the Fresnel method is capable of simulating field distribution with high precision. In contrast, the largest disk results presented previously with conventional boundary element method is about $50~\um$ in diameter~\cite{Zou:11}, $100$ times smaller than the structure presented in this article.
\begin{figure}[htbp]
\subfloat[]{
\label{fig:ARCR6umWavelengthQ}
\begin{minipage}[b]{\linewidth}
  \includegraphics[height=0.75\linewidth,width=1\linewidth]{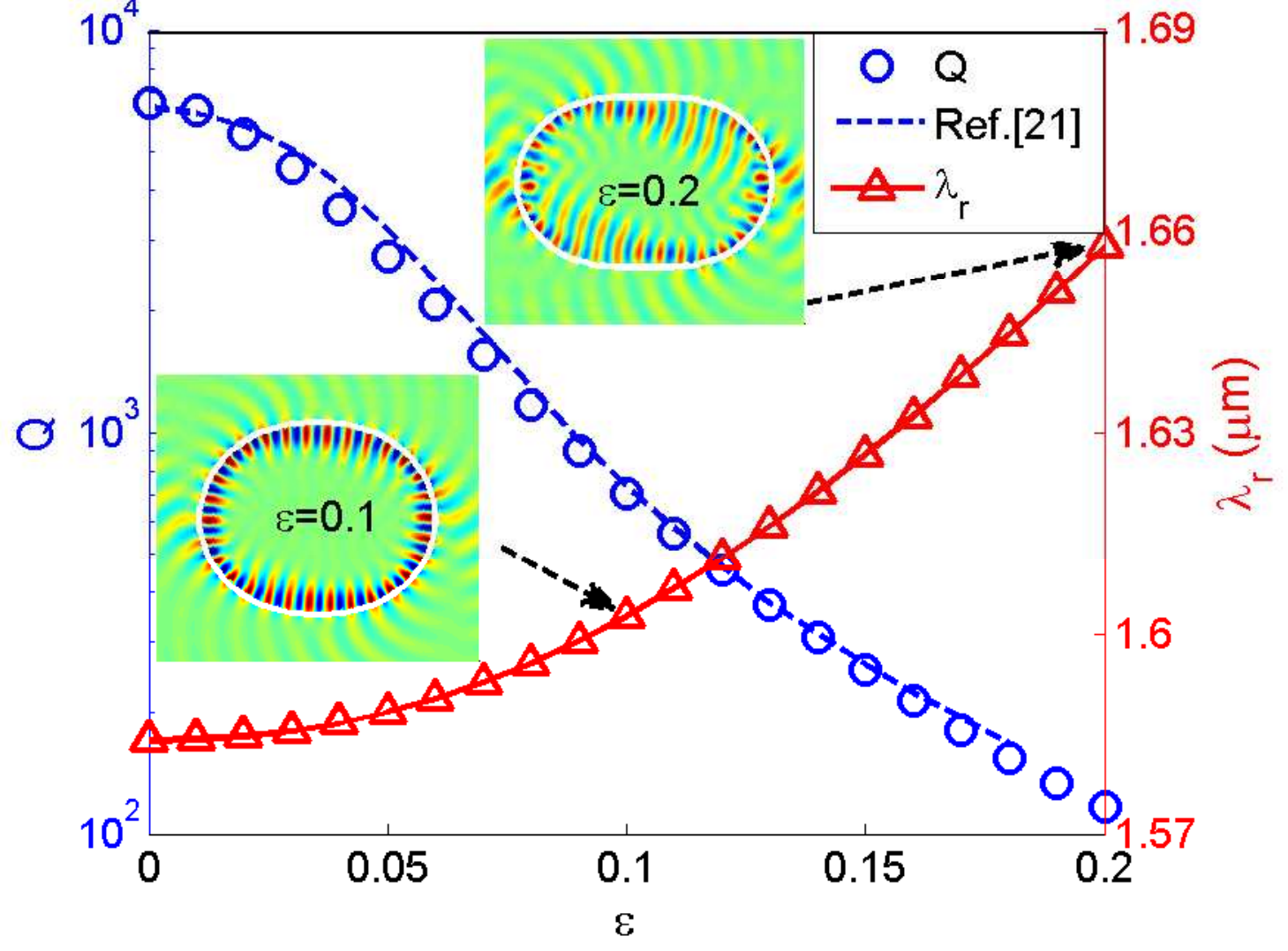}
\end{minipage}}
\vspace{-0.15em}
\subfloat[]{
\label{fig:DoubleDisksCouplingR50umGauss16SeriesNE128}
\begin{minipage}[b]{\linewidth}
  \includegraphics[width=0.9\linewidth]{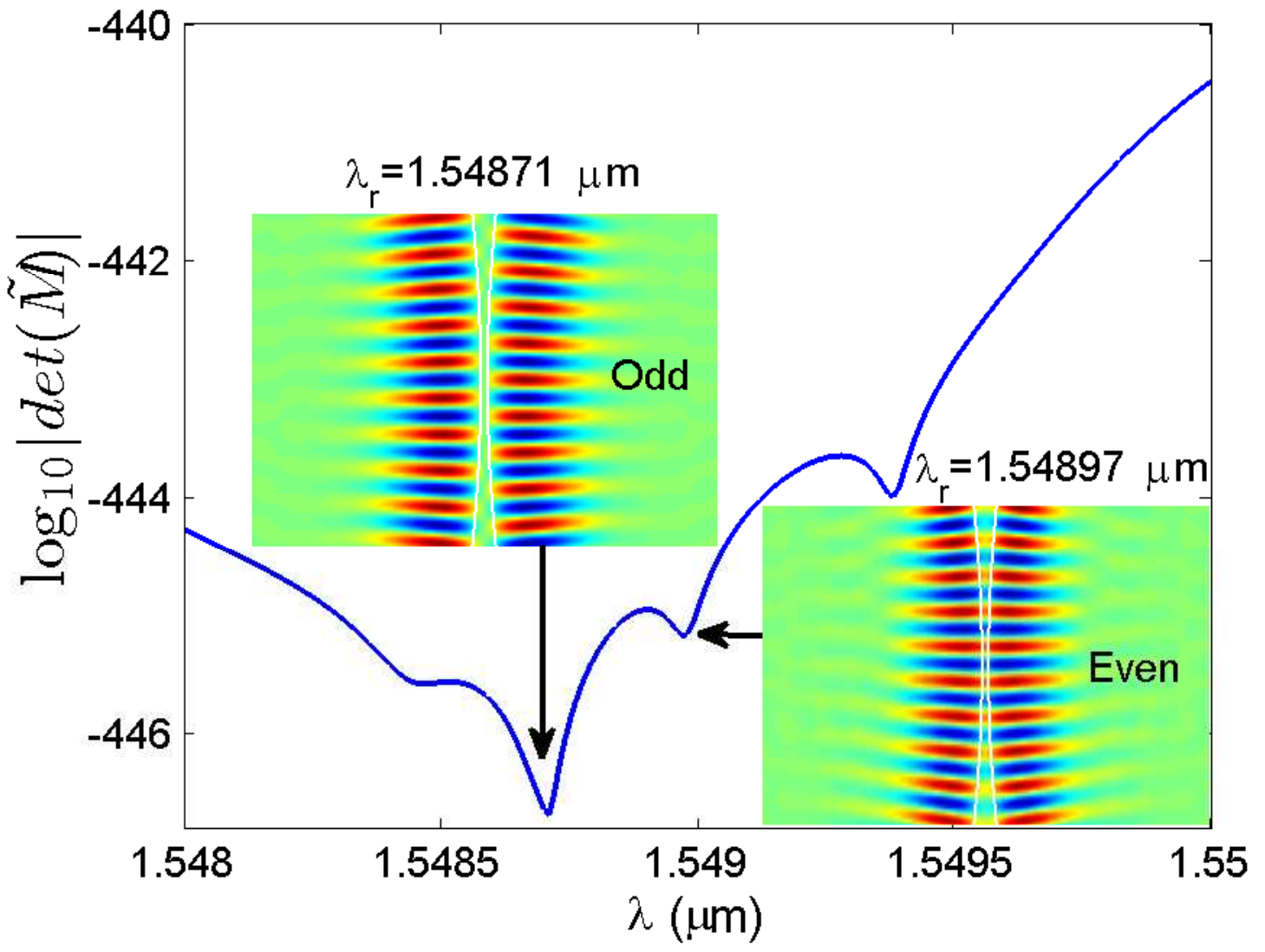}
\end{minipage}}
 \caption{(a) The quality factor and resonance wavelength of quadrupole shaped microcavities as a function of the deformation factor $\varepsilon$. The dashed line is obtained from Ref. [21] as a reference. The insets are the field distribution when $\varepsilon=0.1$ (left) and $\varepsilon=0.2$ (right). (b) The plot of the magnitude of $\mathbf{\tilde M}$ determinant of two microdisks spaced at a $0.2~\um$ gap as a function of wavelength reveals two hybrid modes at resonance wavelengths of $1.54871~\um$ and $1.54897~\um$, corresponding to an even and an odd parity mode as evident by the field distribution plots (insets).}
 \label{fig:Applications}
\end{figure}

Finally we extend our simulations to non-ideal whispering gallery cavities.  We first modelled quadrupole shaped microcavities~\cite{Nockel1996Directional} whose boundary coordinates  are defined as $\rho(\phi)=R(1+\varepsilon\cos(2\phi))$ in a cylindrical system $(\rho,\phi)$. Here $\varepsilon$ is the deformation factor.  The resonance wavelengths (red triangles) and quality factor (blue circles) as a function of the deformation factor are presented in Fig.~\ref{fig:ARCR6umWavelengthQ}. As a comparison, the previously published quality factor~\cite{Zou2009Accurately} of the same structure is plotted as a blue dashed line. As shown, the two set of results are in good agreement. Note that at $\varepsilon=0$, we obtain a $Q=6.61{\times}10^3$, accurate to the third digits compared to the analytical value of $Q=6.60{\times}10^3$~\cite{Ryu2008Resonances}. In Ref.~\cite{Zou2009Accurately}, the corresponding $Q$ yields a slightly larger error of $6.43{\times}10^3$. As indicated from the intensity distribution, at $\varepsilon=0.1$ (left inset) the directional emission is less evident compared to the disk with $\varepsilon=0.2$ displayed in the right inset.
We further investigated a pair of identical microdisks spaced at a gap distance of $0.2~\um$. Here, the diameters of both disks are set to $100~\um$, a five fold increase compared to previously reported modelling of $20~\um$-diameter double disks~\cite{Schwefel2009improved,Grudinin2009Thermal,Grudinin2012Finite}. In Fig.~\ref{fig:DoubleDisksCouplingR50umGauss16SeriesNE128}, two minimum determinants are located at wavelengths of $1.54871~\um$ and $1.54897~\um$. The field distribution plot displayed as insets confirms that the corresponding hybrid modes are of even and odd parity respectively. Note in our calculation, we set the boundary element size to be $1.23~\um$, corresponding to a total of $512$ boundary elements used in the computation.

\section{Conclusion}
In conclusion, by implementing Fresnel technique into boundary element method, the efficiency can be improved by orders of magnitude without additional cost of inaccuracy. As a result, larger cavity structure beyond the capacity of conventional boundary element method can be modelled. In addition, the improved accuracy enables one to compute quality factors of an ultra-high Q cavity with direct root searching techniques.


\end{document}